\documentclass{aa}  
\usepackage{graphicx}
\usepackage{txfonts}
\usepackage{aalongtable,lscape}
\usepackage{natbib}
\bibpunct{(}{)}{;}{a}{}{,}

\begin{document}
   \title{Imaging of star clusters in unperturbed spiral galaxies with the Advanced Camera for Surveys\thanks{Based on 
       observations made with the NASA/ESA Hubble Space Telescope, obtained at the Space Telescope Science Institute, 
       which is operated by the Association of Universities for Research in Astronomy, Inc., under NASA contract 
       NAS 5-26555. These observations are associated with program \# 9774.}}

   \subtitle{I. The low luminosity galaxy NGC~45}
   \authorrunning{M.\ D.\ Mora et al.}
   \titlerunning{The low luminosity galaxy NGC~45}

\author {M. D. Mora\inst{1} \and
         S.S. Larsen\inst{2} \and
	 M. Kissler-Patig\inst{1}}
 \offprints{M.\ D.\ Mora}

   \institute{European Southern Observatory, Karl-Schwarzschild-Strasse 2,85748 Garching bei Munich.Germany
     \and
     Astronomical Institute, University of Utrecht, Princetonplein 5,
     NL-3584 CC, Utrecht, The Netherlands \\
     \email{mmora@eso.org, larsen@astro.uu.nl, mkissler@eso.org}
             }

   \date{Received ???; accepted ???}

 
  \abstract
   {Star clusters are present in almost all types of galaxies. Here we 
    investigate the star cluster population in the low-luminosity, unperturbed 
    spiral galaxy NGC~45, which is located in the nearby Sculptor group. Both
    the old (globular) and young star-cluster populations are studied.}
   {
    Previous ground-based observations have suggested that NGC~45 has few if
    any ``massive'' young star clusters.  We aim to study the population of 
    lower-mass ``open'' star clusters and also identify old globular clusters 
    that could not be distinguished from foreground stars in the 
    ground-based data.
   }
 %
   {
    Star clusters were identified using $UBVI$ imaging from the
    \emph{Advanced Camera for Surveys (ACS)} and the \emph{Wide Field Planetary 
    Camera 2 (WFPC2)} on board the \emph{Hubble Space Telescope}.
    From  broad band  colors and comparison with simple stellar population (SSP) models 
    assuming a fixed metallicity, we   derived the age, mass, and extinction. 
    We also measured the radius for each star cluster candidate.
     }
   {
     We identified 28 young star cluster candidates. 
     While the exact values of age, mass, and  
     extinction depend somewhat on the choice of SSP models, we find no
     young clusters with masses higher than a few 1000 $M_{\odot}$ for
     any model choice.  We derive the luminosity function of young star 
     clusters and find a slope of $\alpha=-1.94 \pm0.28 $. We also identified 19   
     old globular clusters, which appear to have a mass distribution that is 
     roughly consistent with what is observed in other globular cluster systems.
     Applying corrections for spatial incompleteness, we estimate a specific 
     frequency of globular clusters of  $S_N$=1.4--1.9, which 
     is significantly higher than observed for other late-type galaxies 
     (e.g.\ SMC, LMC, M33). Most of these globular clusters appear to belong to a
     metal-poor population, although they coincide spatially with the
     location of the bulge of NGC~45.
   }
  {}


   \maketitle
%

\section{Introduction}

Especially since the launch of the HST, young star clusters have been observed 
in an increasing variety of environments and galaxies.  This includes 
interacting galaxies such as NGC~1275 \citep[e.g.][]{holtzman1992},
the Antennae system \citep[e.g.][]{whitmore2}, tidal tails 
\citep[e.g.][]{nate2005}, but also some normal disk galaxies \citep[e.g.][]{soeren}.
This shows that star clusters are common objects that can form in all
star-forming galaxies.
It remains unclear what types of events trigger star cluster formation 
and the formation of ``massive'' clusters in particular.  It has been 
suggested that (at least some)  globular clusters may have been formed in
galaxy mergers \citep{schweizer}, 
 and the observation of young massive star clusters in the Antennae and elsewhere may be
an important hint that this is indeed a viable mechanism, although not
necessarily the only one.
In the case of normal spiral galaxies, spiral arms may also stimulate the 
molecular cloud formation \citep{elmegreen1994} and thus the 
possibility of star cluster formation.

While a large fraction of stars appear to be forming in clusters initially,
many of these clusters ( $\sim$90\%) will not remain bound after gas
removal and disperse after $\sim10^7$ years \citep{whitmore}.
This early cluster disruption may be further aided by mass loss due to the 
stellar evolution and dynamical processes \citep{fall}, so that many
stars initially born in clusters eventually end up in the field.

Much attention has focused on star clusters in extreme environments such
as mergers and starbursts, but
little is currently known about star and star cluster formation in 
more quiescent galaxies, such as
low-luminosity spiral galaxies.  The Sculptor group  is the nearest galaxy 
group, and it hosts a number of late-type galaxies with luminosities similar to 
those of SMC, LMC, and M33 \citep{cote}. One of the outlying 
members is NGC 45, a low surface-brightness spiral galaxy with M$_B=-17.13$  
and distance modulus $(m-M)=28.42 \pm 0.41$ 
\citep{bottinelli}. This galaxy was included in the
ground-based survey of young massive clusters (YMCs) in nearby spirals
of \citet{lr99}, who found only one cluster candidate.
Several additional old globular cluster candidates from ground-based observations were found by 
\citet{knut}, but none of them has been confirmed.

In this paper we aim at studying star cluster formation in this galaxy using the 
advantages of the HST space observations.  We identify star clusters  
through their sizes,  which are expected to be stable in the lifetime 
of the cluster \citep{spitzer}. Then we derive their ages 
and masses using broadband colors with the limitations that this method 
implies, such as models dependences \citep{degrijs}.
Also we study how the choice of model metallicities affects our results.

This paper is structured in the following way, beginning in Sect. 2, we describe 
the observations, reductions, photometry, aperture corrections
and, artificial object experiments. In Sect. 3, we  describe the 
selection of our cluster candidates, the color magnitude diagram and their spatial distribution.
 In Sect. 4 we describe the properties of young star clusters. In Sect. 5 we
 comment on the globular cluster properties, and Sect. 6 contains the
discussion and conclusions.

\section {Data and reductions}

\begin{figure}
   \centering
   \includegraphics[width=8cm]{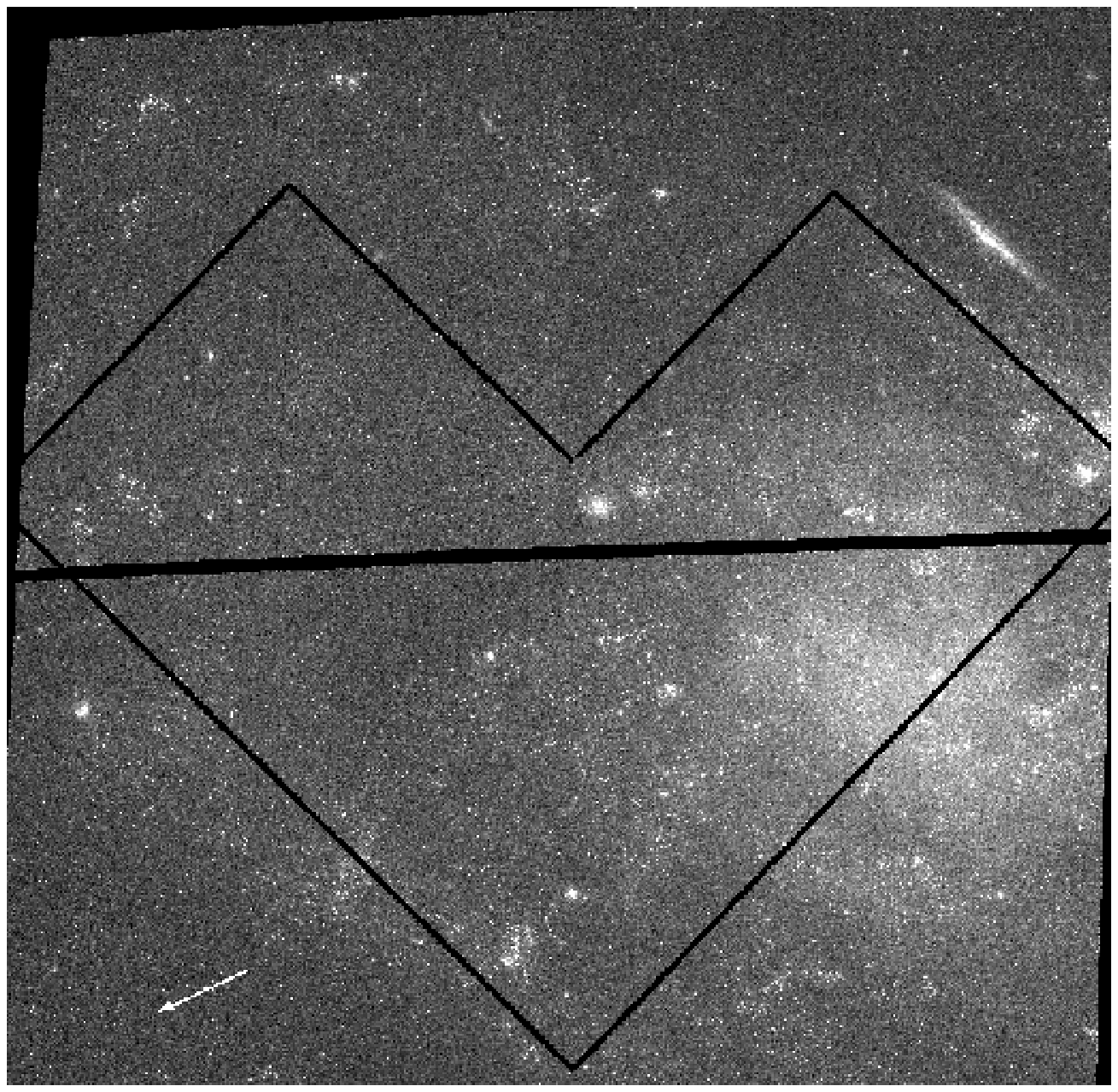}
   \includegraphics[width=8cm]{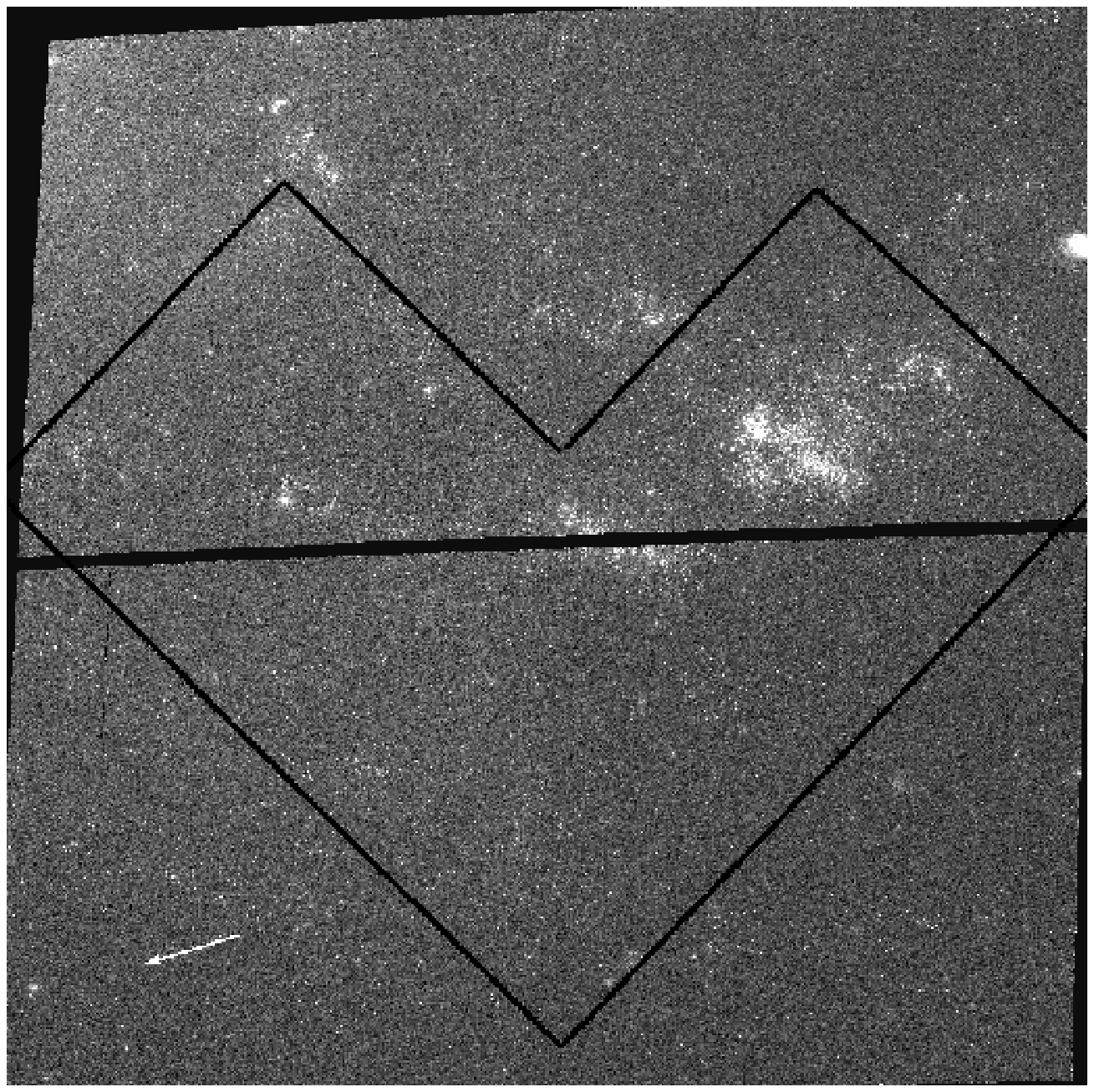}
      \caption
	  {
	    The two ACS images of NGC 45 with the HST WFPC2 (F336W) pointings
	    also indicated.  The arrows indicate the north.  
	  }
          \label{pointing}
  \end{figure}

  Two different regions in NGC~45 were observed with the HST ACS Wide Field 
Channel on July 5, 2004.  One pointing included the center of the galaxy 
($\alpha_{2000}=00^{h} 14^{m} 0^{s}.30$, 
 $\delta_{2000}=-23^{\circ} 10\arcmin 04\arcsec$) 
and the other covered one of the spiral arms 
($\alpha_{2000}=00^{h} 14^{m} 14^{s}.90$,  
 $\delta_{2000}=-23^{\circ} 12\arcmin 29\arcsec$). 
For each frame,  two exposures of 340 seconds each were acquired through the 
filters F435W ($\sim B$) and F555W ($\sim V$), and a pair of 90 and 340 
seconds was obtained through the filter F814W ($\sim I$). In addition, for each pointing, two
F336W ($\sim U$-band) exposures of 1200 s each were  taken with the 
WFPC2. The ACS images are shown in Fig.~\ref{pointing}, including the
footprint of the WFPC2 exposures. Due to the smaller field-of-view of
WFPC2, only part of the ACS frames have corresponding $U$-band imaging.

Following standard ``on-the-fly'' pipeline processing, the raw ACS images 
were drizzled using the MULTIDRIZZLE task \citep{multidrizzle}  in the
STSDAS package in IRAF\footnote{IRAF is
distributed by the National Optical Astronomical Observatory, which is
 operated by the Association of Universities for Research in Astronomy, 
Inc, under  cooperative agreement with the National Science Foundation.}. 
For most of the parameters in Multidrizzle we used the default values.
However, we disabled the automatic sky subtraction, because did not work
well for our data, due to the highly non-uniform background level.
The WPFC2 images were combined using the CRREJ task and standard 
parameter settings.

\subsection{Photometry}

The source detection was carried out in the ACS $F435W$ images using 
SExtractor V2.4.3 \citep{BERTIN}.  
The object coordinates measured in the $B_{F435W}$ frame were used as input
for the SExtractor runs on the other two ACS frames. An area of 5 connected
pixels, all of them are more than 4 sigma above the background, was defined
as an object.  From the output of SExtractor we kept the FWHM measured in 
each filter and the object coordinates. 

Aperture photometry was done with the PHOT task in IRAF, using the SExtractor 
coordinates as input.  This was preferred over the SExtractor magnitudes
because of the greater flexibility in DAOPHOT for choosing the background
subtraction windows.  We used an aperture radius of 6 pixels for the
ACS photometry, which matches the typical sizes of star clusters well at the well  
distance of NGC~45 (1 ACS/WFC pixel $\sim$ 1.2 pc).  The sky was subtracted 
using an annulus with inner radius of 8 pixels and a 5 pixel width. 

Because the $U_{F336W}$ exposures were not as deep as the ACS exposures, we used 
the ACS object coordinates transformed into the WFPC2 frame in order to 
maximize the number of objects for which $U_{F336W}$ photometry was available.
We defined a transformation 
between the ACS and WFPC2 coordinate systems using the GEOMAP task in IRAF, and transformed
the ACS object lists to the WFPC2 frame  
with the GEOXYTRAN task. Each transformation typically had an rms of 
0.5 pixels.  The transformed coordinates were used as input for the WFPC2 
aperture photometry.  We used a 3 pixel aperture radius, which is the 
same physical aperture as in the ACS frames.
The counts were converted to the Vega magnitude system using the
HST zero-points taken form the HST web page\footnote{http://www.stsci.edu/hst/acs/analysis/zeropoints/} and  based on the spectrophotometric calibration of Vega from \citet{bohlin}.

We identified  a total  of 14 objects in common with the ACS data and the WFPC2 
chips. For the Planetary Camera (PC), we were not able to find any 
common objects to determine the correct transformation.  
  \begin{figure}
   \centering
   \includegraphics[width=8cm]{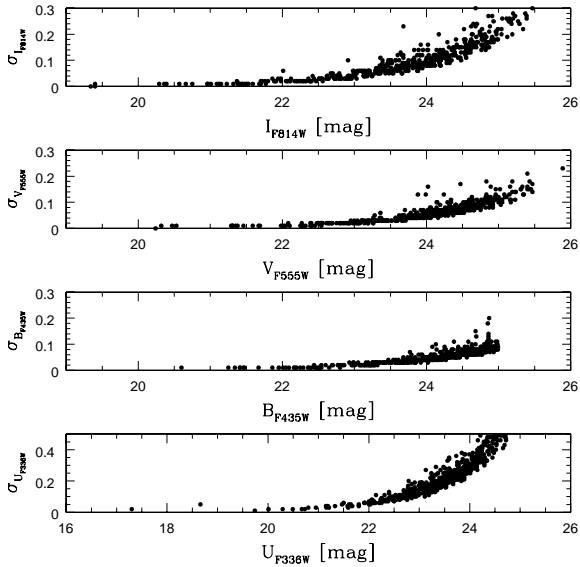}
      \caption{Photometry errors of the detected sources. The aperture radii
              used for the photometry are 6 pixels for the ACS data and 
	      3 pixels for WFPC2.
              }
         \label{photometryerr}
   \end{figure}
\subsection{Object sizes}

Measuring object sizes is an important step in  disentangling stars 
from extended objects. We performed size measurements on the ACS data. 
 For this purpose we used the ISHAPE task in BAOLAB 
\citep{Larsen99}.  ISHAPE models a source as an analytical function 
(in our case,  \citealt{king} profiles) convolved  with the PSF.
For each object Ishape starts from  an initial value for the FWHM, 
ellipticity, orientation, amplitude, and object position, which are then used 
in an $\chi^{2}$  iterative minimization. The output includes the derived 
FWHM, chi-square, flux, and signal-to-noise for each object plus a residual  
image. 
A King concentration parameter of c=30, fitting radius of 10 pixels, and a 
maximum centering radius of 3 pixels were adopted as input parameters for  Ishape. 
These results are described in  Sect. \ref{selection}.

\subsection{Aperture corrections}

Aperture corrections from our photometric apertures to a reference
($1\farcs45$)
 aperture should ideally be derived using the same objects 
in all the frames\footnote{Globular clusters' half light radii are in the range 
1-10 pc; i.e. clusters will appear extended in ACS images at the distance of NGC~45}. 
Those objects should be extended, isolated, and easily 
detectable.  In our case this was impossible because most of the objects were 
not isolated, and the few isolated ones were too faint in the WFPC2 frame. 
For this reason, we decided to create artificial extended objects and derive 
the aperture corrections from them. We proceed for the ACS as follows.

First, we generated an empirical PSF for point sources in the ACS images, 
using the PSF task in DAOPHOT running within IRAF.  Since we want to be sure 
that we are only selecting stars in the PSF construction,
we used ACS images of the Galactic globular cluster 47 Tuc
\footnote {Based on data obtained from the 
ESO/ST-ECF Science Archive Facility.}. We selected  139, 84, and 79 stars 
through the  $B_{F435W}$, $V_{F555W}$, and $I_{F814W}$ filters with 10 sec, 150 sec, and 
72 sec of exposure time, respectively. We could not use the same objects in 
all the filters because the ACS frames have different pointings and exposure 
times.  The selected stars were  more or less uniformly distributed over 
the CCD, but we avoided the core of the globular cluster because stars there 
were crowded and saturated. We used a PSF radius of 11 pixels and a fitting 
radius of 4 pixels on each image. This was the maximum possible radius for 
each star without being affected by the neighboring one.

Second, models of extended sources were generated using the baolab 
\emph{MKCMPPSF}  task \citep{Larsen99}. This task creates a PSF by 
convolving a user-supplied profile (in our case the empirical ACS PSF) with an 
analytical profile (here a \citealt{king}  model with concentration 
parameter $r_{\rm tidal}/r_{\rm core} = 30$) with a FWHM specified by the 
user. The result is a new  PSF for extended objects. 

Third,  we used the \emph{MKSYNTH} task in BAOLAB to create an artificial 
image with artificial extended sources on it. 

For the WFPC2 images we proceeded in a similar way, but  we used the Tiny Tim \citep{Krist} 
package  for the PSF generation.  
We kept the same PSF diameter as for the ACS images. We then followed the 
same procedure as for the ACS PSF.  Aperture corrections were done taken into account the profile used 
for size measurements (King30) and the size derived from it for each object.
Since sizes were measured in the ACS frames, we assumed that objects in the
WFPC frames have similar sizes. Aperture corrections were corrected from  6 pixels for the ACS  data and 3 pixels for WFPC2 to
a nominal $1\farcs45$ (where aperture corrections start to remain constant) reference aperture.
 The corrections are listed  in Table \ref{Aperture} 

\begin{table}
\caption[]{Aperture corrections as a function of object size (FWHM).}
\label{Aperture}
$$
\begin{array}{ccccc}
  \hline
  \hline
  \noalign{\smallskip}
  (1)&(2)&(3)&(4)&(5)\\
  FWHM (pix) & U_{F336W} [mag] & B_{F435W} [mag] & V_{F555W} [mag] & I_{F814W} [mag] \\
             & WFPC2           &      ACS        &     ACS         &        ACS      \\
  \hline
  0.20-0.75  & -0.050 & -0.053 & -0.046 & -0.054  \\ 
  0.75-1.50  & -0.168 & -0.167 & -0.153 & -0.163  \\
  1.50-2.15  & -0.358 & -0.343 & -0.337 & -0.345  \\
  2.15-2.75  & -0.517 & -0.506 & -0.491 & -0.505  \\
  \noalign{\smallskip}
  \hline
\end{array}
$$
\end{table}

Colors do not change significantly as a function of size.
From Table \ref{Aperture}, we note that an error in size will be
translated into a magnitude error: e.g. 0.3 pixel of error in the measured FWHM 
correspond to $\Delta m=0.07$ which, translated into mass,
  corresponds to a $7\%$ error. We also keep in mind that adopting an average correction over each  small size ranges
(0.20-0.75, 0.75-1.50, 1.50-2.15, 2.15-2.75) and introduce an additional uncertainty in mass of $\sim 8\%$

The other systematic effect on our measurements is that
aperture correction changes for the different profiles. We adopt a King30 profile, fitting  most of our objects best. 
For comparison we give some examples  of the effect below. 
For an object  of FWHM=0.5 pixels, the  aperture correction   varies from 
$\Delta m=-0.036$ [mag] ($3\%$ in  mass)  
considering a KING5 profile up to $\Delta m =-0.205$ [mag] ($17\%$ in mass) considering a King100 
profile.  For an object of FWHM=1.2 pixels, the aperture correction   varies 
from $\Delta m =-0.045$ [mag] using King5 ($4\%$ in mass),  up to $\Delta m =-0.437$ [mag]
($35\%$ in mass) considering a King100 profile. And for an object of FWHM=2.5 pixels, 
the aperture correction varies from $\Delta m =-0.078$ ($7\%$ in mass) considering a  King5 profile up 
to $\Delta m =-0.645$ [mag] ($ 45\% $ in mass) considering a King100 profile.
Thus in general the exact size and assumed profiles will cause errors  in mass.

\subsection {Artificial object experiments}
\begin{figure}
   \centering
   \includegraphics[width=8cm]{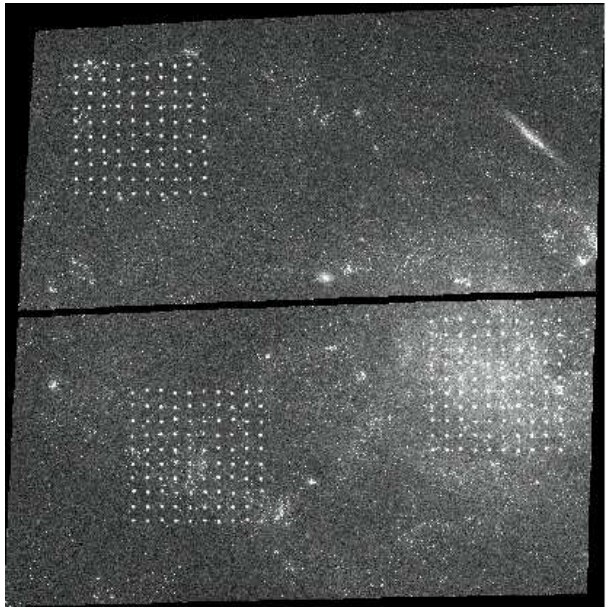}
        \caption
	    {
	      Artificial objects added to the science image. The three 
	      different test  fields are shown, starting from the
	      right and going clockwise: Field I (high background, and crowded 
	      background), Field II (young star-forming region), 
	      Field III (low background region). 
	    }
            \label{artificial}
  \end{figure}

We need to estimate the limits of  our sample's reliability  in magnitudes and sizes.
In the following we investigate how factors such as the degree of crowding and the
background level affect the detection  process and size derivation. To
do so, we added 100 artificial objects and repeated the analysis 
for 3 different sub-regions in the ACS images: ``field I'' was centered
on the bulge of the galaxy, ``field II'' included a crowded region with many 
young stars, and ``field III' covered a low-background region far from the 
center of the galaxy (see Fig. \ref{artificial}).
Each field measured 1000 x 1000 pixels, and the artificial objects were 
distributed in an array of 10 by 10.  A random shift between 0 and  20 pixels 
was added to the original object positions. 
In this way, the minimum separation between objects is  60 pixels.

The artificial objects were built using an artificial PSF as described in the second 
step of the aperture correction.
Objects with a fixed magnitude were then added to a zero-background image 
(as in the third step of aperture correction).
Finally we added this image, containing the artificial objects, to the 
science image using the IMARITH task in IRAF.  This was done for objects with 
magnitudes between  m($B_{F435W}$)=16 and m($B_{F435W}$)=26 and different  FWHM 
(0.1 pixels (stars), 0.5, 0.9, 1.2, 1.5, and 1.8 pixels).
The recovering process was performed  in exactly  the same  way as for the original science 
object detections
(SExtractor detection, aperture photometry, and Ishape run).
  \begin{figure}
   \centering
   \includegraphics[width=8cm]{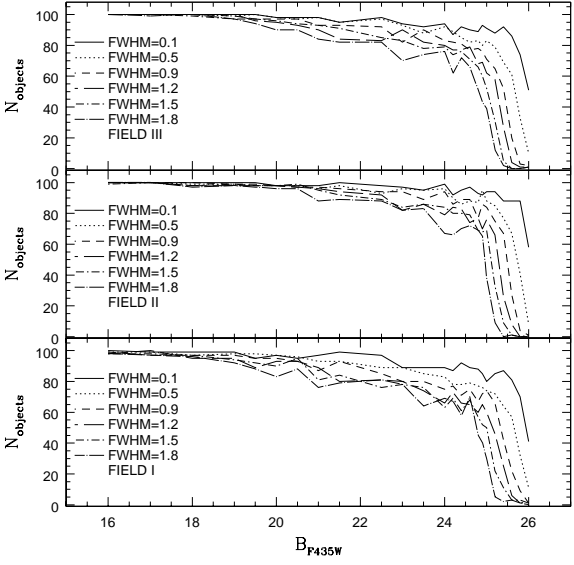}
        \caption
	    {
	      Completeness profiles for each of the test ``fields''  in the 
	      ACS image. The different lines represent different FWHMs.
	    }
        \label{complemag}
   \end{figure}
Figure~\ref{complemag} shows the fraction of artificial objects recovered
as a function of magnitude for different intrinsic object sizes and for
each of the three test fields. As expected, the completeness tests show
that more extended objects are more difficult to detect at a fixed magnitude.
Fields II and III show very similar behavior, perhaps not surprisingly,
since the crowded parts of field II cover only a small fraction of the test
field. The higher background level in Field I results in a somewhat shallower
detection limit, but for all the fields the 50\% completeness limit is reached
between  m($B_{F435W}$)=25 and m($B_{F435W}$)=26.
  \begin{figure}
   \centering
   \includegraphics[width=8cm]{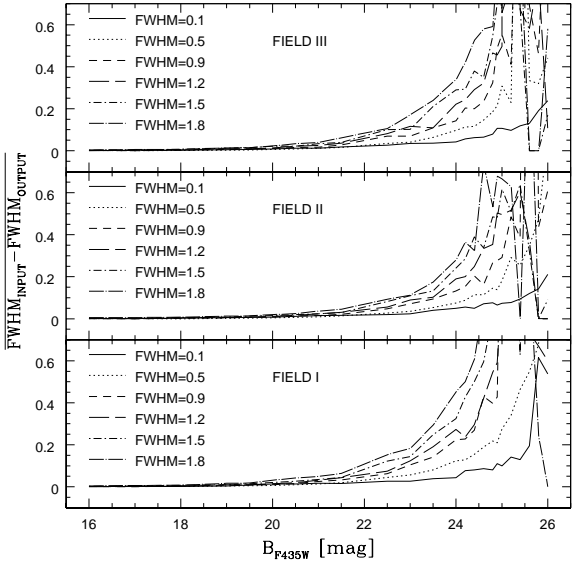}
        \caption
	    {
	      Absolute value of the average of the input FWHM and the recovered value 
	      for each magnitude on each field.  Faintest magnitudes show 
	      biggest FWHMs differences. The 0 $\sigma$ observed in the 
	      fields II and III at $B_{F435W}\sim25.5$ means that no objects
	      were recovered in these bins.  
	    }
            \label{errorFWHM}
  \end{figure}
%

The artificial object experiments also allow us to test the  reliability of 
the size measurements.  In Fig. ~\ref{errorFWHM} we plot the average value 
of the absolute difference between the input FWHM and the recovered
FWHM as a function of magnitude for the three different fields. More extended 
clusters show bigger absolute differences between the input and output 
FWHM at fixed magnitude compared with the less extended. 
Uncertainties are generally larger in the crowded and high background regions. 
In Field 1, the artificial object tests give an average difference between 
the input FWHM and the recovered one of $\sigma$=$0.006$ pixels for
an object with m($B_{F435W}$)$=$20 and FWHM(input)=0.5 pixels. The corresponding
errors at m($B_{F435W}$)=23 and m($B_{F435W}$)=24 are $\sigma$=$0.06$ pixels and
$\sigma$=$0.12$ pixels, and a 50\% error ($\sigma$=$0.25$ pixels) is reached at 
m($B_{F435W}$)$\sim$25. The relative errors also remain roughly constant at a fixed
magnitude for more extended objects; for FWHM=$1.8$ pix the 50\% error
limit is at m($B_{F435W}$)=24.6.


\section {Selection of cluster candidates}   
\label{selection}

For the selection of star cluster candidates, we took advantage of the
excellent spatial resolution of the ACS images. At the distance of NGC~45,
one ACS pixel ($0\farcs05$) corresponds to a linear scale of about 
1.2 pc. With typical half-light radii of a few pc 
\citep[e.g.][]{soeren},
young star clusters are thus expected to be easily recognizable as extended
objects. However, the high spatial resolution and depth of the ACS images
also add a number of complications: as discussed in the
previous section, our detection limit is at $B\sim25.5$, or $M_B\sim-3$ at the distance of NGC~45.
Clusters of such low luminosities are often dominated by a few bright
stars, so it can be difficult to distinguish between a true star cluster
and chance alignments of individual field stars along the line-of-sight.
Therefore, we  limit ourselves to brighter objects for which reliable sizes can be measured.
We adopt a  magnitude limit of m($B_{F435W}$)=23.2,  
corresponding to $M_{B} \sim 5.5$ and  to a size uncertainty of $\sim20$\% for
objects with FWHM $\leq 2.5$ pixel. Only one object (ID=9) has a larger FWHM than
this.

\begin{figure}
  \centering
  \includegraphics[width=8cm]{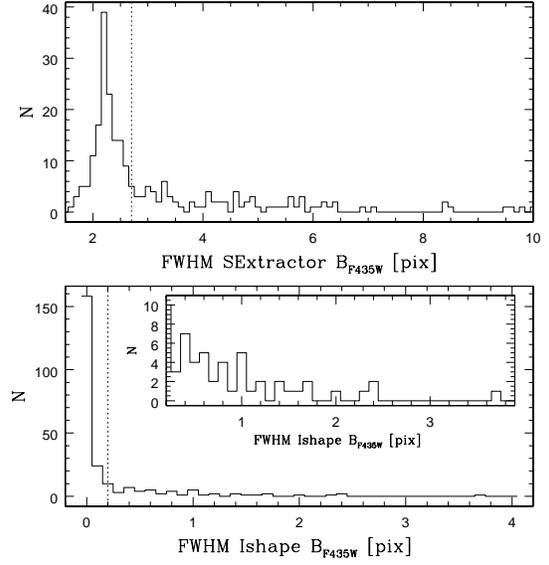}
  \caption
      {
	Histogram of the size distributions from ISHAPE with a zoom in to
	the extended objects (bottom) and
	SExtractor (top). The dotted line shows the size selection criteria. 
      }
      \label{FWHM}
\end{figure}

In an attempt to improve our object detection scheme, we use both 
size estimates from ISHAPE and  SExtractor.  
In Fig. \ref{FWHM} we plot the distribution of size measurements 
for all the objects with $B_{F435W}\le23.2$.

\begin{figure}
  \centering
  \includegraphics[width=8cm]{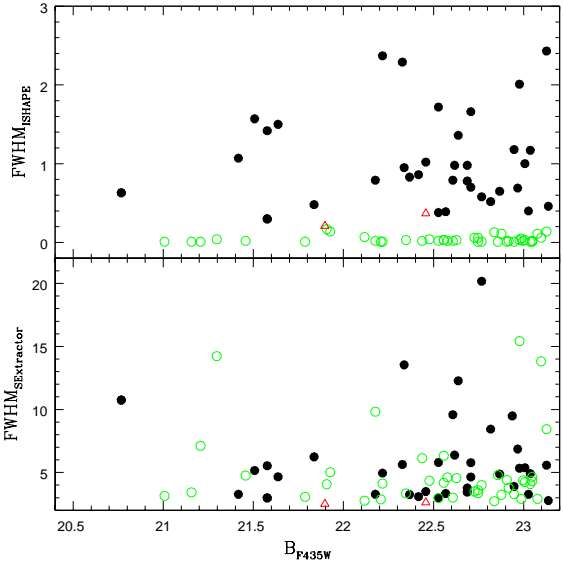}
  \caption
      {
        Object FWHMs as measured by ISHAPE and SExtractor. Filled circles are 
        objects considered extended by both methods. Open circles were 
        considered as extended objects only by SExtractor and  triangles 
        are objects considered extended only by ISHAPE.
      }
      \label{FWHMvsMAG}
\end{figure}

The SExtractor sizes show a peak around FWHM$\sim$2.5, corresponding to the
PSF FWHM, while the ISHAPE distribution peaks at 0 (recall that the ISHAPE sizes
are corrected for the PSF).  We adopt the size criteria of 
$FWHM_{\rm SEx} \ge 2.7$ and  $FWHM_{\rm ISHAPE} \ge 0.2$ to select extended objects.  
In  Fig.  \ref{FWHMvsMAG} we plot all objects with at least one condition fulfilled.
From this figure it is evident that several objects were considered to be extended 
according to the SExtractor sizes, while the ISHAPE fit yield to near-zero
size. Such objects might be close groupings of stars where SExtractor 
yields the size of the whole group, while ISHAPE fits a single star. 
On the other hand, there are very few objects that are classified as
extended by ISHAPE but as compact according to SExtractor.

Objects that fulfill both conditions
are considered to be star cluster candidates.  Objects that fulfill
only one condition  are considered to be possible clusters. Objects 
that did not pass any condition were rejected (i.e.\ classified as stars). 
All the objects were visually inspected  to avoid contamination by HII regions. 
Finally 66 objects were rated as star cluster candidates and  64 as possible clusters,
while 59 of these ``possible'' star cluster were considered as stars by ISHAPE and
2 as stars by SExtractor.
From the 66 star cluster candidates, 36 have 4-band photometry and 30 have
only ACS (3-band)  photometry.
In the following, only star cluster candidates are considered for the analysis.
%
%
%
\subsection{Young clusters vs globular clusters}
\begin{figure}
   \centering
   \includegraphics[width=8cm]{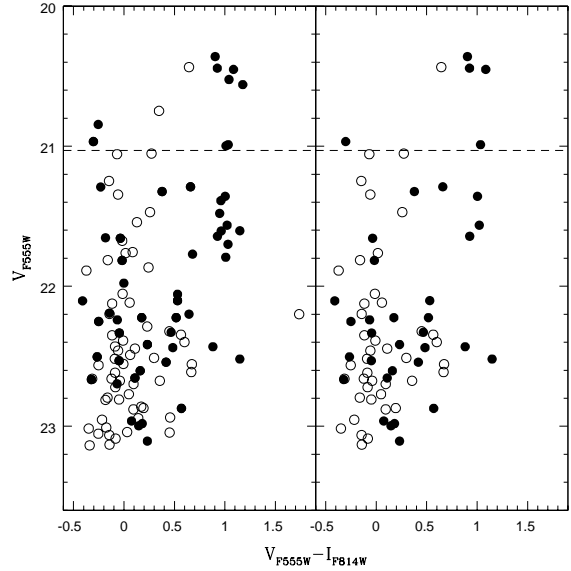}
   \caption
       {
	 Color-magnitude diagram for objects in NGC~45. The left side
	 corresponds to all objects with $B_{F435W}$,$V_{F555W}$ and, 
	 $I_{F814W}$ ACS photometry and  the right side correspond to objects with 
	 4-band photometry (ACS filters plus $U_{F336W}$ from WFPC2).
	 Filled circles  are extended objects (i.e. star clusters) selected
	 by SExtractor and Ishape, while open circles are objects 
	 that did not pass one of the size selection criteria. 
	 The dashed line  is the TO of the old MW globular clusters system
	 M$_{V,TO} \sim-7.4$.
       }
   \label{HR}
\end{figure}

In Fig.~\ref{HR} the color magnitude diagrams are shown for star clusters candidates and possible clusters.
Two populations can be distinguished: A population of blue (and probably young) 
objects with $V_{F555W}-I_{F814W}\le 0.8$ (with a main concentration around 
$V_{F555W}-I_{F814W}\sim0$), and a red population  $V_{F555W}-I_{F814W}\ge0.8$, concentrated
around $V_{F555W}-I_{F814W}\sim1$ (globular clusters). Two blue clusters brighter than $V_{F555W}=21$
($M_V\sim-7.5$) are found. The red objects have colors consistent with
those expected for old globular clusters and  are all  extended according
to both the SExtractor and ISHAPE size criteria. 

  \begin{figure}
   \centering
   \includegraphics[width=8cm]{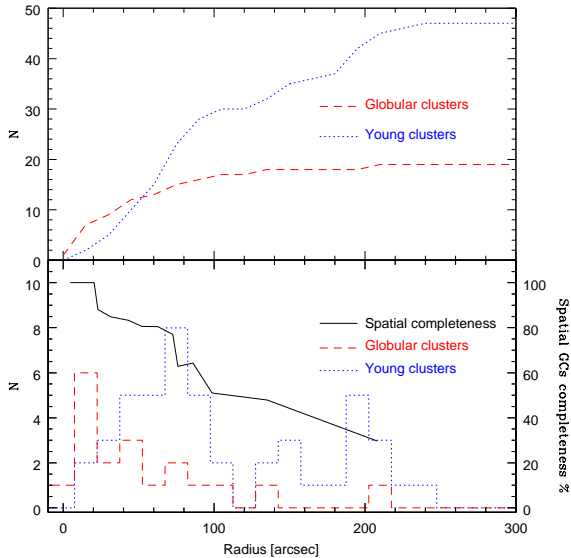}
      \caption
	  {Star cluster spatial distribution from the center of the galaxy at different radius.
	    Top: Cumulative distribution with radius.
	    Bottom: Number of objects per radius and globular cluster spatial completeness. } 
	  \label{DISTRIBUCION}
  \end{figure}

The spatial distribution is shown in Fig. \ref{DISTRIBUCION} for all the clusters with 3-band 
photometry (ACS field of view). 
Globular clusters are concentrated towards the center of the galaxy, and their number 
decreases with their distance. In contrast, young clusters are distributed in the outer 
part of the galaxy showing 2  major concentrations at 78 arcsec and 190 arcsec.
Young clusters are associated with star forming regions, therefore it is
 more likely to find them in spiral arms like the first concentration rather than in the center. 
The second concentration corresponds mainly to the clusters detected in the second pointing covering one of 
the spirals arms and young regions.

Spatial completeness is $100\%$  up to 20 arcsec radius.  Beyond 20 arcsec, the completeness drops down to $20\%$ at 200 arcsec radius. For all completeness corrections, we assume the 
non-covered areas to be similar to the covered area in all respects.

The detection of  young and globular star clusters do not vary differentially with radius. 
Therefore the central concentration of the globular clusters is not an artifact of detection completion.

Assuming that the covered area is representative of
the non-covered area and that the number and distribution of star clusters are similar in the non-covered area, we
expect that the observed  tendency of globular clusters being located towards the center of the galaxy and the young
ones toward the outer parts will  remain.

In the following we discuss the young clusters  and globular clusters  separately. 
Our sample includes 36 star clusters with UBVI data (28 young star clusters and 8 globular clusters) and
11 globular clusters with BVI data. We will 
not discuss  further the ages nor masses derived for the globular clusters since they are unreliable
due to their faint $U_{F336W}$ band magnitudes. 
Only for completeness we show their derived ages and masses in the table \ref{tableGC4}.

\section{Young star clusters}

In this section, we discuss the properties of young clusters in more detail.

\subsection{Colors}

  \begin{figure}
   \centering
   \includegraphics[width=8cm]{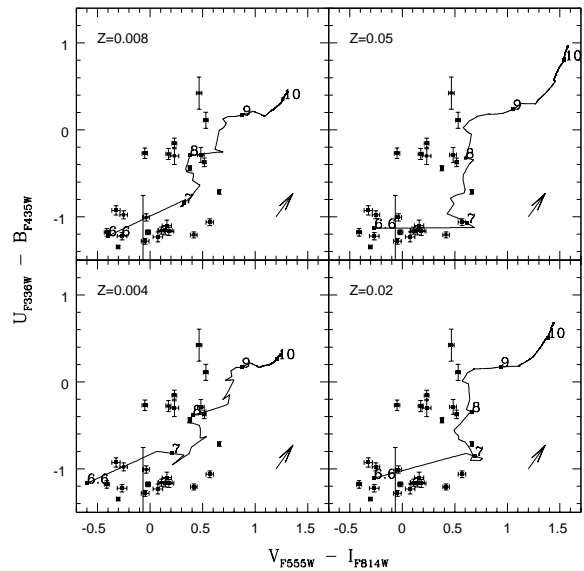}
      \caption
	  {
	    Color-color diagram of NGC~45 star cluster candidates, uncorrected
	    for reddening.
	    Each line corresponds to SSP GALEV models of different 
	    metallicities for a Salpeter IMF and ages from 1 Myr up to 10 Gyr.
	    The arrow corresponds to an extinction of 1 mag.  }
         \label{colornored}
   \end{figure}
The $U_{F336W}-B_{F435W}$ vs.\ $V_{F555W}-I_{F814W}$ two-color diagram of NGC~45 young star cluster
candidates (not reddening corrected) is shown in Fig. \ref{colornored}.
Also shown in the figure are GALEV \citep{anders} SSP models
for different metallicities. Age increases along the tracks from blue
towards redder colors. The ``hook'' at $V_{F555W}-I_{F814W}$$\sim0.5$ and
$U_{F336W}-B_{F435W}$$\sim-1$ corresponds to the appearance of red supergiant stars
at $\sim10^7$ years and is strongly metallicity dependent.
Many of the  clusters have colors consistent with very young ages
($<10^7$ years), and some older cluster candidates are spread along the
theoretical tracks.

The cluster colors show a considerable scatter compared with the model
predictions, significantly larger than the photometric errors.
For younger objects, the scatter may be due to 
random fluctuations in the number and magnitude of  red supergiant stars 
present in each cluster \citep{girardi}. Reddening
variations can also contribute to the scatter, and for the older 
clusters ($\ga$ 1 Gyr), metallicity effects may also play a role since 
the models in each panel follow a fixed metallicity.

Generally, the cluster colors seem to be better described by models of
sub-solar metallicity (Z=0.004 and Z=0.008). Considering that the
luminosity of NGC~45 is intermediate between those of the small and 
large Magellanic clouds, we might indeed expect young clusters to
have  intermediate metallicities between those typical of young stellar 
populations in the Magellanic clouds.

  \begin{figure}
   \centering
   \includegraphics[width=8cm]{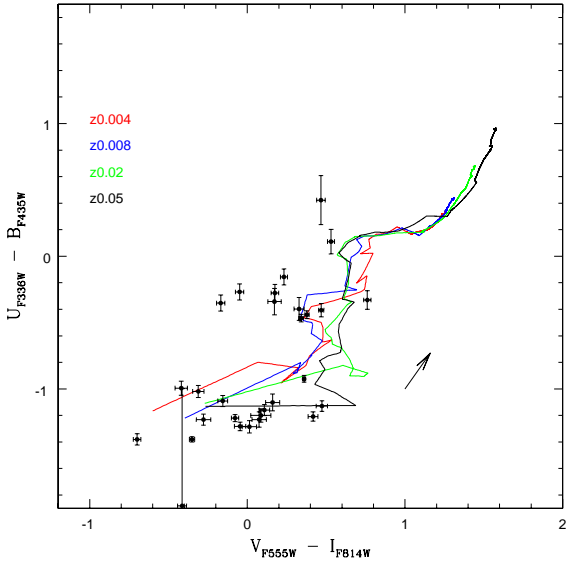}
      \caption
	  {
	    Reddening-corrected color-color diagram of NGC~45 star clusters. 
	    Each cluster was corrected according to the lowest $\chi^2$ metallicity-fitted model.
	    All the theoretical tracks are plotted for 4 different metallicities. 
	    See Tables \ref{tableYoung} and \ref{tableGC4} for the derived values.
	  }
         \label{colorred}
   \end{figure}

%
%
\subsection{Ages and masses}
\label{agesection}
One of main problems in deriving ages, metallicities, and masses for star 
clusters in spirals is  that we do not know the extinction towards 
the individual objects. \citet{arjan} propose a method known 
as the ``3D fitting'' method to solve this problem.  This method estimates 
the extinction, age, and mass by assuming a fixed metallicity for each single cluster.
 The method relies on a SSP model 
(in our case GALEV assuming a Salpeter IMF,  \citealt{anders}), 
which provides the broad-band colors as a function of age and metallicity. 
The algorithm compares the model colors with the observed ones and searches 
for the best-fitting extinction (using a step of 0.01 in $E(B-V)$) and age 
for each cluster, using a minimum $\chi^{2}$ criterion.  Finally the mass 
is estimated by comparing the mass-to-light ratios predicted by the models 
(for a fixed metallicity) with the observed magnitudes.

In Fig. \ref{colorred} we plot the model tracks for 4 metallicities, together with
the cluster colors corrected for reddening, according to the best 
$\chi^2$ fitting for the different metallicities.
 The 3D fitting method will move the clusters in the opposite direction with respect to the reddening 
arrow in Fig. \ref{colorred}, finding the closest matching model. 

From Hyperleda \citep{paturel}, the internal extinction in B-band for
NGC 45 is $A_B=0.34$, based on the inclination and the morphological galaxy type taken from \cite{Bottinelli95}. 
Assuming $A_B=1.324*A_V$ from \cite{rieke}, we  expect $E(B-V)\sim0.08$. 
The mean derived extinction for the clusters is $E(B-V)= 0.04$ for $Z=0.008$ and $Z=0.004$,
$E(B-V)= 0.05$ for $Z=0.02$, and 
$E(B-V)= 0.1$ for $Z=0.05$. These values agree well with our derived value. Considering that some 
clusters lie in the foreground and some in the background, the extinction value from the literature
agrees well with the extinctions we derive for our star clusters.

\begin{figure}
   \centering
   \includegraphics[width=8cm]{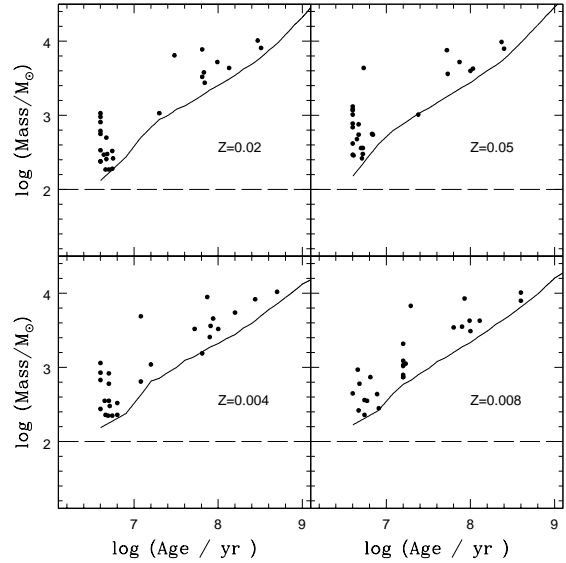}
   \caption
	  {
	    Mass as a function of cluster age. All star cluster candidates
	    with masses greater than 100 solar masses are plotted here.
	    Filled circles are  young clusters. The lines represent a cluster
	    of $F435W=23.1$ at different ages and masses 
	    for each metallicity. 
	  }
         \label{massage}
   \end{figure}

The 3D fitting method was applied for all clusters with 4-band photometry  
assuming in turn for different metallicities (Z=0.004, Z=0.008, Z=0.02, and Z=0.05).  
To explore how the assumed metallicity affects the derived parameters,
we plot the cluster masses in Fig.  \ref{massage}  against cluster ages for 
all the models.
In a general overview of each plot we  see concentrations of clusters
around particular ages, e.g. near $\log(Age/yr) \sim7.2$ and $\sim8$
in the Z=0.008 plot. These concentrations are  not physical, but instead
artifacts due to the model fitting (see \citealt{arjan}). In 
particular, the concentration around $log(Age/yr) \sim 7$ is due to the rapid 
change in the integrated cluster colors at that age (corresponding to the 
``hook'' in 
Fig.~\ref{colornored}). The figure again illustrates that the exact age at 
which this feature appears is  metallicity dependent.

Independent of metallicity, Fig.~\ref{massage} shows a concentration of 
young and not very massive clusters ($M \la 10^3 M_\odot$) around
$10^{6.8}$ yr. 
At older ages, the number of detected cluster candidates per age bin
(note the logarithmic age scale) decreases rapidly.
This is a result of fading due to stellar evolution (as indicated by 
the solid lines), as well as cluster disruption (see Sect.~\ref{sec:disrupt}). 

\subsection{Sizes}

  The young cluster candidates detected in NGC~45 are generally low-mass, 
compact objects.  The average size (from ISHAPE) is FWHM=1.16 $\pm$ 0.2 pixel, 
equivalent to a half-light radius of R$_{eff}$= 2.0 $\pm$ 0.2 pc (errors are the 
standard error of the mean). These mean sizes are somewhat smaller than
those derived by \citet{soeren} for clusters in  a sample of normal spiral galaxies, 
which typically range from 3--5 pc. 
Previous work on star clusters has shown that there is at most a shallow 
correlation between cluster sizes and masses.  For example, young star 
clusters found in NGC~3256 by \citet{zepf} and by \citet{soeren} in
a sample of nearby spiral galaxies show a slight correlation between their 
radii and masses, although
\citet{nate2005b} do not find any apparent relation between 
radii and masses in M51.
\citet{soeren} derived the following relation between cluster
mass and size: $R_{\rm eff} = 1.12 \times (M/M_{\odot})^{0.10}$ pc. For a cluster
mass of $10^3$ M$_{\odot}$, which is more typical of the young clusters 
observed here, this corresponds to a mean size of 2.2 pc. Thus, the
difference between the cluster sizes derived by other studies of extragalactic
star clusters and those found for NGC~45 here may be at least partly due
to the lower cluster masses encountered in NGC~45. Of course, biases in
the literature studies also need to be carefully considered. For example,
\citet{soeren} excludes the most compact objects, which might cause the
mean sizes to be systematically overestimated in that study.

\begin{figure}
  \center
   \includegraphics[width=8cm]{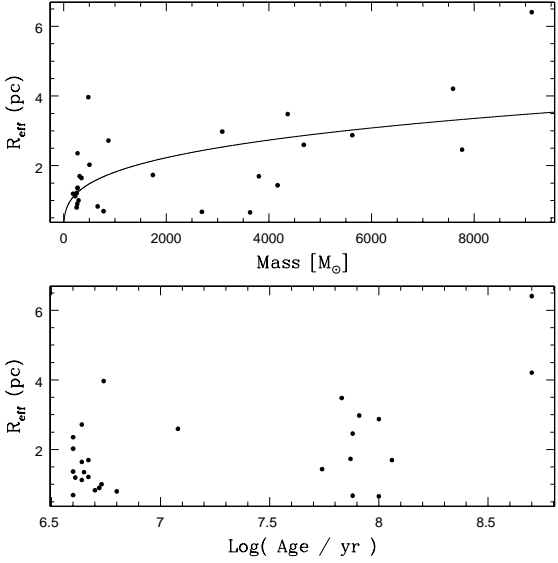}
      \caption
	  {
 	   Effective radius vs mass (top) and  age (bottom). The curve on top 
 	   is the fitting of a power-law $R_{eff}=a*M^b$
	  }
         \label{relac}
   \end{figure}

Figure \ref{relac} shows the effective radius versus mass and age for 
the cluster candidates in NGC~45 (mass estimates are for Z=0.008). The 
plotted line is the best-fitting relation of the form :
\begin{equation}
  R_{eff}=a*M^b
\end{equation}
where $a=0.24$ $\pm$ $0.16$ and $b=0.29 \pm 0.08$ are the best-fitting values. 
As in previous studies, there is a slight tendency for the more massive
clusters to have larger sizes, but our fit  has a large scatter and the 
constants are not tightly constrained. Therefore we suggest that the
NGC~45 sample cannot be taken as  strong evidence of a size-mass relation,
but it does not contradict the general shallow trends found in other
studies.

\subsection{Cluster disruption time}                                                                %

\label{sec:disrupt}

\citet{BoutloukosLamers} defined the 
disruption time as
\begin{equation}
t_{dis}(M)=t_{4}^{dis}(M/10^4M_{\odot})^{\gamma},
\end{equation}
where $t_{4}^{dis}$ is the disruption time of a cluster with an initial mass 
of $10^4 M_{\odot}$. The constant $\gamma$ has been found empirically
to have a value of about $\gamma=0.6$ \citep{BoutloukosLamers}.
If a constant number of clusters are formed per 
unit time (constant cluster formation rate) and clusters are formed in 
a certain mass range with a fixed cluster initial mass function (CIMF),
which can be written as a power law
\begin{equation}
N(M)\sim M^{-\alpha},
\end{equation}
then the number of clusters (per age interval) detected above a certain 
fixed magnitude limit depends only on fading due to stellar evolution,
as long as there is no cluster disruption.  When cluster disruption becomes
significant, this behavior is broken and the number of clusters decreases 
more rapidly with time. In this simple scenario, no distinction is made 
between cluster disruption due to various effects (interaction with the 
interstellar medium, bulge/disk shocks, internal events such as two-body 
relaxation), and it is assumed that a single ``disruption time-scale'' 
applies (with the mass dependency given above).

\begin{figure}
   \centering
   \includegraphics[width=8cm]{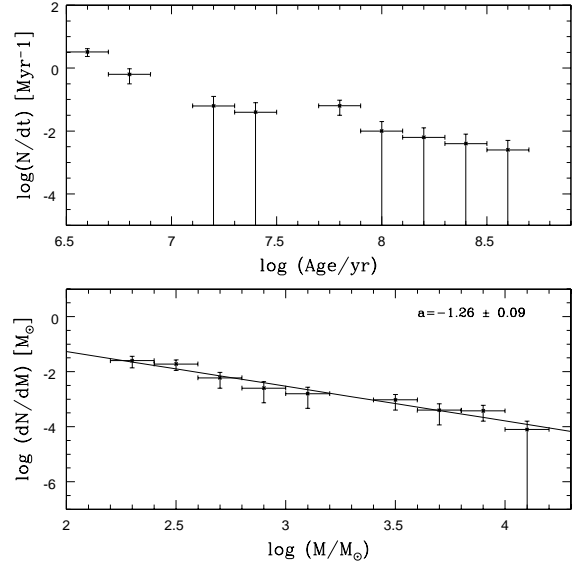}
      \caption
	  {
           Age distribution (top) and mass distribution of clusters (bottom). The line
	   is the fitting line of slope $a=-1.26\pm0.09$. 
	  }
         \label{Cross}
   \end{figure}

Under these assumptions,
the timescale on which cluster disruption is important can be derived
from the cluster age- and mass distributions, which are both expected
to show a break:
\begin{equation}
  \log \left(\frac{t_{\rm cross}}{10^8} \right) = 
    \frac{1}{1-\gamma\zeta} \left[ \log 
    \left(\frac{t_4^{\rm dis}}{10^8}\right)
     + 0.4 \gamma (m_{\rm ref} - B_{\rm lim})
    \right]
\end{equation}
\begin{equation}
  \log \left(\frac{M_{\rm cross}}{10^4} \right) = 
    \frac{1}{1-\gamma\zeta} \left[ \zeta \log 
    \left(\frac{t_4^{\rm dis}}{10^8}\right)
     + 0.4 (m_{\rm ref} - B_{\rm lim})
    \right]
\end{equation}
(Eqs.\ 15 and 16 in \citealt{BoutloukosLamers}). In
these equations, $V_{\rm lim}$ is the detection limit, and $m_{\rm ref}$
is the apparent magnitude of a cluster with an initial mass of
$10^4 M_{\odot}$ at an age of $10^8$ years, the subscript ``cross'' is the  breaking  point between the cluster fading and the cluster disruption 
and $\zeta$ gives the rate of fading due to stellar evolution.

Figure~\ref{Cross} shows the age- and mass distributions for young cluster 
candidates in NGC~45.  
Figure~\ref{Cross} shows no obvious break in either the mass- or
age distributions. In order to see a hint of a break, it would be  necessary to 
include the lowest-mass and/or youngest bins, but as discussed above, the
age determinations are highly uncertain below $10^7$ yr, as is the
identification of cluster candidates with masses of only $\sim100 M_{\odot}$.
Considering that the sample is $\sim$ 80\% complete at m($F435W$)$=23.2$,
it is unlikely that completeness effects can be responsible for 
the lack of a break in the mass- and age distributions.
Furthermore, many of the objects with ages below $\sim10^7$ years may
be unbound and not ``real'' star clusters. We may thus consider $10^7$ years
as an upper limit for any break and therefore $t_{cross}\la 10^7$ years.
Similarly, we may put an upper limit of $\log M_{\rm cross} \la 2.5$ 
on any break in the mass function. 
Nevertheless, considering the slope value and assuming  $\alpha=2.0$, we derived
 $\gamma=0.73 \pm 0.09$ that agrees with the  $\gamma$ value found in other studies. 
  \begin{figure}
   \centering
   \includegraphics[width=8cm]{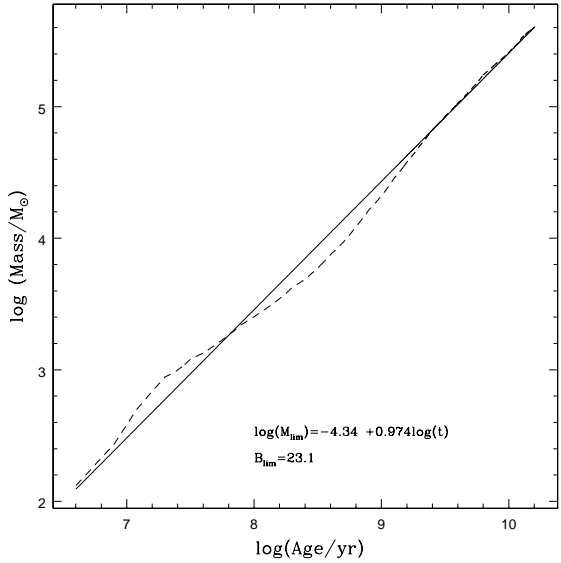}
   \caption
          {
            Detection  limit for all ages and masses for a cluster of $B_{F435W}=23.1$ 
            represented by the dashed line.  The solid line is a linear fit
	      to this relation.
          }
         \label{fitages}
   \end{figure}

In Fig. \ref{fitages} we plot the detection limit (in mass) vs.\ age for 
$B_{lim}=23.1$. From this we derived $\zeta=0.97$ and $0.4(m_{ref}-B_{lim})=-0.782$. Fixing
$\gamma=0.6$ and using Eq. 15 from \citet{BoutloukosLamers}, we  get
$\log(t_4 / 10^8) \la 0.05$
and finally :
\begin{equation}
\log{t_{dis}} \la 8 + 0.6\log{M_{cl}/10^4}
\end{equation}
where $t_{dis}$ is in years and mass is in $M_{\odot}$. Since the young
clusters in NGC~45 generally have masses below $10^4$ M$_{\odot}$, the
disruption times will accordingly be less than $10^8$ years, and it is therefore
not surprising that no  large number of older clusters are observed
(apart from the old globular clusters).


\subsection{Luminosity function}

\begin{figure}
  \centering
  \includegraphics[width=10cm]{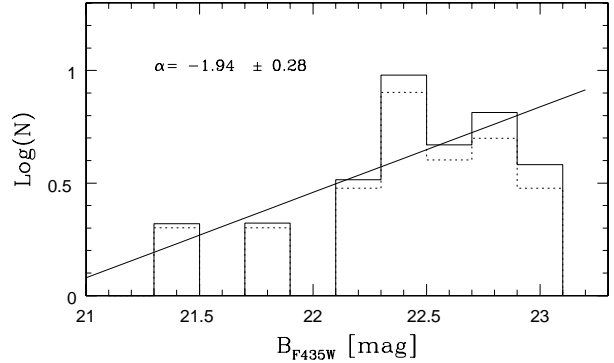}

  \caption
      {
	$B_{F435W}$ band luminosity function of the cluster candidates.
	The dashed histogram is the uncorrected luminosity function, while
	the solid histogram is the completeness-corrected one.  
	The solid line represents the power-law fit of the form dN(L)/dL $\alpha$ L$^{\alpha}$ 
	to the corrected LF for a cluster FWHM=1.2 pixels.
      }
      \label{Lfun}
\end{figure}

Figure \ref{Lfun} shows the luminosity function of young star clusters in NGC~45,
the luminosity function without correction for incompleteness, i.e.
for all the clusters with  $B_{F435W}<23.2$, and the completeness-corrected luminosity 
function  for a source with FWHM=1.2 pixels. 
The histogram was fitted using $\chi^2$ with a relation of the form 
\begin{equation}
\centering
\log N = a M_B + b,
\end{equation}
which yield $a=0.37 \pm 0.11$, $b=-7.9 \pm 2.5$. This can be converted to the more common
representation of the luminosity function as a power-law 
$dN(L_{B})/dL_{B}={\beta}{L{_{B}}}^{\alpha}$,
using Eq. 4 from \citet{soeren2}:
\begin{equation}
\centering
\alpha=-(2.5a+1),
\end{equation}
which yields $\alpha=-1.94 \pm0.28$.

~This value is in agreement (within the errors)  with  slopes found in \citet{soeren2} and other studies of  \textbf{$-2.4$ $\leq$ $\alpha$ $\leq$ $-2.0$} for a variety of galaxies. 
~It is also consistent with the LMC value $\alpha=-2.01 \pm0.08$ from Table 5 in \citet{soeren2}.  

Alternatively, the slope of the luminosity function may be  estimated by carrying out a 
maximum-likelihood fit directly to the  data points, thus avoiding binning effects. Such a fit 
is sensitive  to the luminosity range over which the power law is normalized,  however. A fit 
restricted to the luminosity range of the clusters  included in our sample 
($21.17 < B_{F435W} < 23.08$) yields $ \alpha=-1.99\pm0.40$, in good agreement with the fit 
in Fig. \ref{Lfun}. If  we restrict the fitting range to objects brighter than $B_{F435W} =22.5$ 
we get a  steeper slope, but with a larger error: $\alpha=-3.2 \pm0.8$. Likewise, allowing for
 a higher  maximum luminosity in the  normalization of the power-law and including a correction 
for  completeness also leads to a steeper slope. In conclusion, the  luminosity function is
 probably  consistent with earlier studies,  but the small number of clusters makes it difficult
 to provide  tight constraints on the LF slope in NGC~45.

%

%
%
%
\section{Globular clusters}
In this section we comment briefly on the globular clusters in NGC 45.

\subsection{ Sizes and color distribution}

For all extended objects with observed  colors $0.8 < (V_{F555W}-I_{F814W}) <1.2$  
(i.e. globular clusters), we measured an average  $FWHM=1.7 \pm 0.4$ pixels from ISHAPE, which is equivalent 
to a half-light radius of $R_{eff}= 2.9 \pm 0.7$ pc (errors are  the standard error of the mean).
These sizes well agree with the average half-light radius         
found by \citet{jordan} in the ACS Virgo Cluster Survey ($R_{eff}=2.7 \pm 0.35$ pc).                 

Due to the  age-metallicity  degeneracy in optical broad-band colors \citep[e.g.][]{worthey}, we cannot 
independently estimate the ages and metallicities of the globular clusters. 
It is still interesting to compare the color distribution of the globular cluster 
candidates in NGC~45 with those observed in other galaxies, (e.g.\ \citealt{1999AJ....118.1526G}; 
\citealt{kundu01}; \citealt{larsenGC}).
 To do this  we transformed the ACS 
magnitudes to standard Bessel magnitudes, following the \citet{sirianni} recipe. 

\begin{figure}
\centering
   \includegraphics[width=8cm]{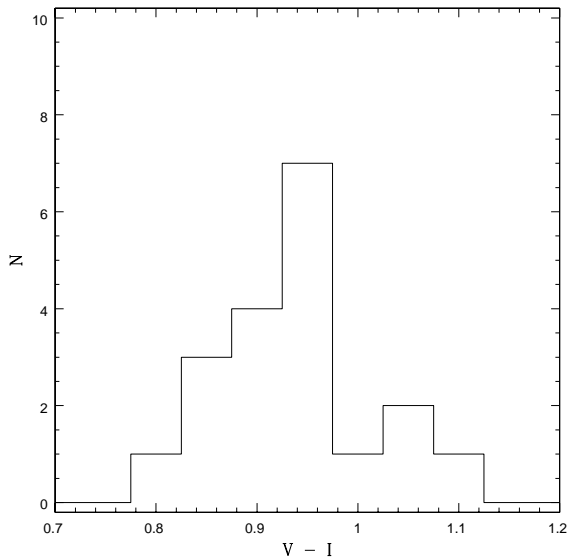}
   \caption
       {
	 Histogram of the $V-I$ color distribution for globular clusters in NGC~45. 
       }
   \label{GCcolor}
\end{figure}

In Fig. \ref{GCcolor} the color $(V-I)$ histogram of the NGC~45 GCs is shown. 
Most of the objects have $V-I \le 1.0 $, whit a mean color of $V-I=0.90 \pm 0.01 $. Three objects 
have redder colors ($(V-I) \ge 1$),  with a mean color of $V-I=1.05 \pm 0.02 $. Thus the majority of 
objects in NGC~45 fall around the blue peak seen in galaxies  where globular clusters exhibit a 
bimodal color distribution. e.g. globular clusters of NGC~45 are very likely metal-poor ``halo'' 
objects \citep[e.g.][]{markus2000}.

%

\subsection{Globular-cluster specific frequency}

\citet{harrisvandenbergh} defined the specific 
frequency S as
\begin{equation}
S=N_{GC}10^{0.4*(M_{V}+15)}
\end{equation}
where $N_{GC}$ is the total number of globular clusters that belong to the 
galaxy and $M_V$ is the absolute visual magnitude of the galaxy. 
The 19 globular cluster candidates detected in NGC~45 provide a lower limit on the globular cluster
specific frequency of 
$S_N = 1.4\pm0.15$, using for NGC~45 $V_{TOT}=10.66\pm0.11$ from the 
HyperLeda\footnote{http://leda.univ-lyon1.fr} database \citep{paturel} and applying a correction for foreground reddening of 
$A_V=0.07$ mag (using $A_B=0.09$ from \citealt{schlegel} and 
assuming $A_B/A_V=1.324$; \citealt{rieke}). 

A more accurate estimate needs to take the detection 
completeness and incomplete spatial coverage into account.  The globular cluster 
luminosity function (GCLF) generally appears to be fit well  by a Gaussian 
 function, and the total number of clusters can therefore be estimated 
by counting the number of clusters brighter than the turn-over and 
multiplying by 2. This reduces the effect of uncertain corrections at the 
faint end of the GCLF (but of course introduces the assumption that the
GCLF does indeed follow a Gaussian shape).
Here we adopted the mean turnover of disk galaxies from Table 14 in 
\citet{carneybook} at 
$M_{TO}=-7.46 \pm 0.08$, i.e. $V=21.03$. This is indicated by the horizontal
line in Fig.~\ref{HR}.

With these assumptions
we calculated $N=13^{+7}_{-2}$  in the observed field of view. 
An inspection of Fig.~\ref{HR} suggests
that our estimate of the GCLF turn-over magnitude may be too
bright, with only 6 (one third) of the detected objects falling above
the line representing the turn-over. This could mean that the distance
is underestimated (note the large uncertainty of $\pm0.41$ mag on
the distance modulus), or that the GCLF does not follow the canonical
shape.

In any case, since our ACS frames cover only part of NGC~45, these numbers
need to be corrected for spatial incompleteness. For this purpose,
we drew a circle centered in the galaxy through each globular cluster, and we counted that
globular cluster as $1/fc$, where $fc$ is the fraction of that circle falling within the 
ACS field of view.  In this way we estimated that 
$N_{exp}=7 \pm3$, and more globular clusters would be expected outside the area
covered by the ACS pointings. Finally, the total estimated number of globular clusters
in NGC~45 is $N_{GC}=20^{+8}_{-3}$ and the specific frequency is
$S=1.5^{+0.8}_{-0.6}$. If instead the 19 clusters are corrected
for spatial incompleteness directly, we estimate a total of 26 globular clusters and
$S_N=1.9 \pm0.7$.

In marked contrast to the relatively modest population of young star
clusters, NGC~45 shows a remarkably large population of old globular clusters
 for its luminosity. We derived  specific frequencies of $S_N=1.9 \pm 0.7$ 
and $S_N = 1.5^{+0.8}_{-0.6}$, depending on 
how the total number of globular clusters is estimated.
This is significantly higher than in other late-type galaxies, e.g.,\ 
\citet{ashmanzepf1998} quote the following values for
late-type galaxies in the Local Group:
LMC ($S_{N}=0.6 \pm 0.2$), M33  ($S_{N}=0.5 \pm 0.2$),
SMC  ($S_{N}=0.2$), and M31  ($S_{N}=0.9 \pm 0.2$). 
Does this make NGC~45 a special galaxy?  Did something happen in the distant 
past of the galaxy?  Globular clusters are distributed in a similar way to the Milky Way halo
globular clusters (i.e. concentrated toward the center).    
The globular cluster color distribution (Fig. \ref{GCcolor}) 
suggests that most of them share the same metallicity, except three that 
might represent a ``metal-rich peak''. Thus, most of the globular clusters in NGC~45 may be
analogues of the ``halo  globular clusters'' in the Milky Way or, more
generally, the metal-poor globular cluster population generally observed in all major
galaxies \citep[e.g.][]{markus2000}.

\section{Discussion and conclusions}

We have analyzed the star cluster population of the nearby, late-type
spiral galaxy NGC~45. Cluster candidates were identified using a size
criterion, taking advantage of the excellent spatial resolution of the
Advanced Camera for Surveys on board HST. In fact, the high resolution
and sensitivity, combined with the modest distance of NGC~45, mean that
the identification of star clusters is no longer limited by our ability
to resolve them as extended objects. Instead, as we are probing farther
down the cluster mass- and luminosity distributions, the challenge is
to disentangle real physical clusters from the random line-of-sight alignments
of field stars. The high resolution also means that the ISHAPE code
may fit individual bright stars in some clusters instead of the total
cluster profile, which can be quite irregular for low-mass objects. Thus,
we rely on a combination of ISHAPE and SExtractor size estimates for the
cluster detection. The detection criteria are conservative, and it is
possible that we may have missed some  clusters.
The ages and masses were then estimated from the
broad-band colors, by comparison with GALEV SSP models.

Our ACS data have revealed two main groups of star clusters in NGC~45. We
found a number of relatively low-mass ($<10^4$ M$_{\odot}$) objects that
have most probably formed more or less continuously over the lifetime of 
NGC~45 in its disk, similar to the open clusters observed in the Milky
Way. We see a high concentration of objects with ages $<10^7$ years,
many of which may not survive as systems of total negative energy.
The mass distribution  of these objects (Fig. \ref{Cross}) is consistent with random
sampling from a power-law. %
The lack of very massive (young) clusters in NGC~45 may then be
explained simply as a size-of-sample effect. We do not see a clear break
in the mass- or age- distributions, and therefore cannot obtain reliable
estimates of cluster disruption times as prescribed by \citet{BoutloukosLamers}. However,
we have tentatively estimated an upper limit of about 100 Myr on the 
disruption time ($t_4$) of a $10^4$ M$_{\odot}$ cluster.
Also, we do not see any evidence of
past episodes of enhanced cluster formation activity in the age distribution
of the star clusters, suggesting that NGC~45 has not been involved in major
interactions in the (recent) past. Thus, star cluster formation in this
galaxy is most likely triggered/stimulated by internal effects, such as
spiral density waves.

Small number statistics, uncertain age estimates, and the difficulty of
reliably identifying low-mass clusters prevent us from determining
accurate cluster disruption time-scales. 
If our estimate of the disruption time is correct, then it is somewhat
puzzling that a large population of old globular clusters are also observed, since 
a $10^5 M_\odot$ cluster should have a disruption time of only $\sim400$ Myrs. 
Given that the globular cluster candidates are usually located closer to the center 
than the young clusters, one might expect an even shorter disruption time
unless disruption of young clusters is dominated by mechanisms that are
not active in the center, such as spiral density waves or encounters
with giant molecular clouds.  
We note that more 
accurate metallicity and age measurements for the globular clusters will require 
follow-up spectroscopy.

\begin{acknowledgements}
We would like to thank Nate Bastian for his useful comments and H.J.G.L.M. Lamers for providing 
the 3D code program. Also we would like to thank the referee Dr. Uta Fritze-von Alvensleben 
for insightful comments that helped in improving this paper.

\end{acknowledgements}


\bibliographystyle{aa}   
\bibliography{moraetal.bib}

\begin{landscape}
\begin{table}
  \caption[]{Young star clusters.}
\tiny
  \label{tableYoung}
  $$ 
  \begin{array}{ccccccccccccccccccccccccccc}
    \hline
    \hline
    \noalign{\smallskip}
 (1)  & (2)     & (3)     & (4)      & (5)       & (6)      & (7)       & (8)       & (9)      & (10)     & (11)      & (12)     & (13)      & (14)     & (15)     & (16)       & (17)       & (18)      & (19)      & (20)           & (21)           & (22)           & (23)           & (24)                & (25)                &(26)                 &        (27)         \\
 ID   & _{RA}   & _{DEC}  &  _{B}    & _{\sigma} &  _{V}    & _{\sigma} &  _{I}    & _{\sigma} &  _{U}    & _{\sigma} & _{FWHMB} & _{FWHMB}  & _{FWHMV} & _{FWHMI} & _{E(B-V)}  & _{E(B-V)}  & _{E(B-V)} & _{E(B-V)} & _{log(Age/yr)} & _{log(Age/yr)} & _{log(Age/yr)} & _{log(Age/yr)} & _{log(M/M_{\odot})} & _{log(M/M_{\odot})} & _{log(M/M_{\odot})} & _{log(M/M_{\odot})} \\
      & _{2000} & _{2000} & _{F435W} & _{F435W}  & _{F555W} &  _{F555W} & _{F814W} &  _{F814W} & _{F336W} &  _{F336W} & _{(pix)} & _{(pix)}  & _{(pix)} & _{(pix)} & _{Z=0.008} & _{Z=0.004} & _{Z=0.02} & _{Z=0.05} & _{Z=0.008}     & _{Z=0.004}     &  _{Z=0.02}     &   _{Z=0.05}    &  _{Z=0.008}         &  _{Z=0.004}         &  _{Z=0.02}          &  _{Z=0.05}          \\

    \noalign{\smallskip}
    \hline
    \noalign{\smallskip}

1  & 0:14:15.46 &  -23:12:48.11 &  21.442 & 0.007 & 21.290 & 0.007 &  20.630 & 0.007 & 20.728 & 0.024 & 4.660  & 1.500 & 1.430 & 0.890   & \textbf{0.25} &  0.21  &  0.07   &  0.41  &  \textbf{7.08}  &   7.29   &   7.48  &  6.73  & \textbf{3.69} & 3.83  & 3.81  & 3.64\\
2  & 0:14:14.95 &  -23:13:17.73 &  22.518 & 0.013 & 22.532 & 0.013 &  22.581 & 0.023 & 22.249 & 0.059 & 3.330  & 0.390 & 0.450 & 0.280   & \textbf{0.00} &  0.10  &  0.33   &  0.35  &  \textbf{7.90}  &   6.89   &   6.60  &  6.60  & \textbf{3.41} & 2.64  & 2.75  & 2.84\\
3  & 0:14:14.92 &  -23:13:23.83 &  22.816 & 0.024 & 22.962 & 0.024 &  22.887 & 0.038 & 21.585 & 0.053 & 4.880  & 0.650 & 0.720 & 1.290   & 0.00 &  0.00  &  0.00  &  \textbf{0.00}  &  6.69  &   7.20   &   6.66  &  \textbf{6.71}   & 2.35 & 2.90  & 2.27  & \textbf{2.42}\\
4  & 0:14:14.70 &  -23:13:20.37 &  22.652 & 0.018 & 22.666 & 0.021 &  22.989 & 0.033 & 21.725 & 0.050 & 5.780  & 0.700 & 0.780 & 0.650   & 0.00 &  0.00  &  0.08  &  \textbf{0.10}  &  6.66  &   6.74   &   6.60  &  \textbf{6.60}   & 2.36 & 2.36  & 2.38  & \textbf{2.47}\\
5  & 0:14:15.40 &  -23:13:55.51 &  22.174 & 0.011 & 22.224 & 0.011 &  21.707 & 0.012 & 21.801 & 0.049 & 3.250  & 0.830 & 0.750 & 0.570   & \textbf{0.04} &  0.02  &  0.00   &  0.00  &  \textbf{7.72}  &   7.80   &   7.83  &  7.73  & \textbf{3.52} & 3.54  & 3.58  & 3.56\\
6  & 0:14:15.00 &  -23:13:50.93 &  22.183 & 0.012 & 22.224 & 0.014 &  22.049 & 0.020 & 21.906 & 0.063 & 5.800  & 1.720 & 1.890 & 1.940   & \textbf{0.00} &  0.00  &  0.00   &  0.41  &  \textbf{7.91}  &   7.90   &   7.81  &  6.60  & \textbf{3.56} & 3.55  & 3.52  & 3.08\\
7  & 0:14:14.17 &  -23:13:20.00 &  22.814 & 0.022 & 23.106 & 0.024 &  22.874 & 0.035 & 22.514 & 0.097 & 5.370  & 1.000 & 0.720 & 1.120   & 0.00 &  \textbf{0.05}  &  0.21   &  0.00  &  7.81  &   \textbf{6.91}   &   6.64  &  7.38  & 3.19 & \textbf{2.45}  & 2.47  & 3.01\\
8  & 0:14:11.63 &  -23:12:18.28 &  22.837 & 0.016 & 22.240 & 0.013 &  22.306 & 0.025 & 21.199 & 0.883 & 4.910  & 1.170 & 1.780 & 2.260   & 0.37 &  \textbf{0.29}  &  0.41   &  0.43  &  6.60  &   \textbf{6.68}   &   6.60  &  6.60  & 2.93 & \textbf{2.78}  & 2.91  & 3.01\\
9  & 0:14:02.07 &  -23:07:58.54 &  22.340 & 0.017 & 22.061 & 0.017 &  21.547 & 0.019 & 24.114 & 0.486 & 9.490  & 3.700 & 4.080 & 3.880   & 0.00 &  0.00  &  \textbf{0.02}   &  0.01  &  8.70  &   8.60   &   \textbf{8.51}  &  8.40  & 4.13 & 4.02  & \textbf{4.05}  & 4.02\\
10 & 0:14:94.67 &  -23:09:22.27 &  22.718 & 0.018 & 22.604 & 0.020 &  22.444 & 0.040 & 21.616 & 0.061 & 20.170 & 0.580 & 0.960 & 1.670   & 0.02 &  0.04  &  \textbf{0.00}   &  0.19  &  6.80  &   7.20   &   \textbf{6.75}  &  6.67  & 2.52 & 3.03  & \textbf{2.42}  & 2.74\\
11 & 0:14:02.66 &  -23:09:33.45 &  22.971 & 0.016 & 22.104 & 0.010 &  22.515 & 0.022 & 21.792 & 0.040 & 3.260  & 0.400 & 0.350 & 0.670   & 0.31 &  \textbf{0.24}  &  0.33   &  0.35  &  6.60  &   \textbf{6.60}   &   6.60  &  6.60  & 2.83 & \textbf{2.65}  & 2.79  & 2.89\\
12 & 0:14:14.03 &  -23:10:00.62 &  21.823 & 0.013 & 21.657 & 0.016 &  21.693 & 0.026 & 20.816 & 0.040 & 5.630  & 2.290 & 3.930 & 2.890   & \textbf{0.10} &  0.03  &  0.22   &  0.24  &  \textbf{6.70}  &   6.81   &   6.60  &  6.60  & \textbf{2.92} & 2.87  & 2.98  & 3.07\\
13 & 0:14:02.55 &  -23:10:32.22 &  22.917 & 0.015 & 22.997 & 0.022 &  22.851 & 0.060 & 21.759 & 0.048 & 6.860  & 0.690 & 0.940 & 2.660   & 0.00 &  0.00  &  0.00   &  \textbf{0.05}  &  6.74  &   7.20   &   6.70  &  \textbf{6.72}  & 2.35 & 2.87  & 2.27  & \textbf{2.48}\\
14 & 0:14:01.45 &  -23:10:05.05 &  22.491 & 0.014 & 22.655 & 0.017 &  22.547 & 0.029 & 21.332 & 0.039 & 3.440  & 0.780 & 0.910 & 1.870   & 0.00 &  0.00  &  0.00   &  \textbf{0.00}  &  6.71  &   7.20   &   6.67  &  \textbf{6.72}  & 2.48 & 3.02  & 2.41  & \textbf{2.56}\\
15 & 0:14:01.45 &  -23:10:04.85 &  22.416 & 0.013 & 22.542 & 0.016 &  22.125 & 0.027 & 21.208 & 0.033 & 6.380  & 0.980 & 0.950 & 3.040   & 0.00 &  0.00  &  0.00   &  \textbf{0.00}  &  7.20  &   7.20   &   6.74  &  \textbf{6.83}  & 3.04 & 3.09  & 2.52  & \textbf{2.75}\\
16 & 0:14:01.15 &  -23:10:21.52 &  22.362 & 0.014 & 22.103 & 0.013 &  21.572 & 0.018 & 22.472 & 0.091 & 4.640  & 1.660 & 1.820 & 2.270   & 0.00 &  0.00  &  \textbf{0.00}   &  0.00  &  8.44  &   8.60   &   \textbf{8.47}  &  8.37  & 3.92 & 4.01  & \textbf{4.01}  & 3.99\\
17 & 0:14:02.37 &  -23:10:59.48 &  23.081 & 0.020 & 22.981 & 0.024 &  22.800 & 0.039 & 21.913 & 0.042 & 2.770  & 0.460 & 0.620 & 0.750   & 0.00 &  0.02  &  0.00   &  \textbf{0.14}  &  6.80  &   7.20   &   6.74  &  \textbf{6.70}  & 2.36 & 2.87  & 2.28  & \textbf{2.56}\\
18 & 0:14:02.83 &  -23:11:10.11 &  22.638 & 0.019 & 22.438 & 0.020 &  21.954 & 0.025 & 22.350 & 0.084 & 5.340  & 2.010 & 1.680 & 1.430   & \textbf{0.13} &  0.10  &  0.00   &  0.00  &  \textbf{7.94}  &   7.99   &   8.13  &  8.03  & \textbf{3.66} & 3.63  & 3.64  & 3.63\\
19 & 0:14:01.65 &  -23:10:58.16 &  21.781 & 0.011 & 21.816 & 0.013 &  21.834 & 0.020 & 20.603 & 0.024 & 6.240  & 0.480 & 0.730 & 0.550   & 0.00 &  0.00  &  0.00   &  \textbf{0.05}  &  6.70  &   7.20   &   6.67  &  \textbf{6.67}  & 2.78 & 3.32  & 2.70  & \textbf{2.88}\\
20 & 0:13:56.63 &  -23:08:48.94 &  22.624 & 0.020 & 22.329 & 0.018 &  21.862 & 0.020 & 23.047 & 0.183 & 5.570  & 2.430 & 2.470 & 2.670   & 0.00 &  0.00  &  \textbf{0.00}   &  0.00  &  8.70  &   8.60   &   \textbf{8.51}  &  8.40  & 4.02 & 3.90  & \textbf{3.91}  & 3.90\\
21 & 0:13:57.52 &  -23:09:15.97 &  22.441 & 0.014 & 22.334 & 0.017 &  22.380 & 0.032 & 21.159 & 0.029 & 12.280 & 1.360 & 1.820 & 2.780   & \textbf{0.00} &  0.00  &  0.00   &  0.09  &  \textbf{6.70}  &   6.77   &   6.68  &  6.65  & \textbf{2.55} & 2.55  & 2.48  & 2.68\\ 
22 & 0:13:56.78 &  -23:08:58.02 &  22.761 & 0.015 & 22.872 & 0.018 &  22.303 & 0.031 & 21.699 & 0.036 & 8.440  & 0.520 & 0.680 & 3.980   & 0.02 &  0.06  &  0.00   &  \textbf{0.08}  &  7.08  &   7.23   &   7.30  &  \textbf{6.84}  & 2.81 & 3.05  & 3.03  & \textbf{2.74}\\
23 & 0:13:59.86 &  -23:10:22.19 &  22.409 & 0.018 & 22.504 & 0.020 &  22.770 & 0.041 & 21.186 & 0.038 & 9.580  & 0.790 & 0.970 & 1.490   & 0.00 &  0.00  &  0.00   &  \textbf{0.01}  &  6.60  &   6.67   &   6.60  &  \textbf{6.61}  & 2.44 & 2.42  & 2.38  & \textbf{2.46}\\
24 & 0:13:59.88 &  -23:10:23.29 &  22.140 & 0.015 & 22.252 & 0.016 &  22.503 & 0.032 & 21.163 & 0.043 & 13.540 & 0.950 & 0.990 & 1.080   & 0.00 &  0.00  &  \textbf{0.05}   &  0.07  &  6.65  &   6.74   &   \textbf{6.60}  &  6.60  & 2.55 & 2.56  & \textbf{2.53}  & 2.62\\
25 & 0:13:59.72 &  -23:10:23.64 &  22.494 & 0.014 & 22.432 & 0.015 &  21.551 & 0.018 & 22.248 & 0.069 & 3.770  & 0.980 & 1.260 & 2.200   & 0.03 &  0.02  &  0.08   &  \textbf{0.10}  &  8.20  &   8.11   &   7.99  &  \textbf{7.87}  & 3.74 & 3.63  & 3.72  & \textbf{3.72}\\
26 & 0:14:01.90 &  -23:11:30.50 &  21.165 & 0.008 & 20.967 & 0.009 &  21.270 & 0.014 & 19.818 & 0.018 & 5.160  & 1.570 & 2.430 & 2.040   & \textbf{0.04} &  0.00  &  0.06   &  0.08  &  \textbf{6.60}  &   6.66   &   6.60  &  6.60  & \textbf{3.06} & 2.97  & 3.03  & 3.12\\
27 & 0:14:00.38 &  -23:10:51.70 &  22.476 & 0.012 & 22.416 & 0.013 &  22.184 & 0.018 & 22.320 & 0.059 & 2.990  & 0.380 & 0.410 & 0.460   & \textbf{0.00} &  0.00  &  0.00   &  0.00  &  \textbf{8.00}  &   8.00   &   7.84  &  8.00  & \textbf{3.52} & 3.49  & 3.44  & 3.60\\
28 & 0:13:59.37 &  -23:10:32.15 &  21.385 & 0.008 & 21.324 & 0.008 &  20.946 & 0.011 & 20.944 & 0.028 & 5.530  & 1.420 & 1.440 & 1.510   & \textbf{0.03} &  0.00  &  0.00   &  0.00  &  \textbf{7.87}  &   7.93   &   7.81  &  7.72  & \textbf{3.95} & 3.93  & 3.89  & 3.88\\
    \noalign{\smallskip}
    \hline
  \end{array}
  $$
{Column (1): Cluster ID. ~Column (2): RA. ~Column (3): DEC. ~Columns (4), (5), (6), (7), (8), (9), (10), (11): Photometry and the 
error in magnitude units. ~Column (12): FWHM derived using SExtractor for the F435W band in pixels. ~Columns (13),(14),(15): FWHM 
derived using ISHAPE in pixels for each band. ~Columns (16), (17), (18), (19): Extinction derived for each metallicity model.  
~Columns (20), (21), (22), (23): Cluster Ages derived for each metallicity model in Log(yrs). ~Columns (24), (25), (26), (27): Cluster 
masses derived for each metallicity model in solar masses. All the values in \textbf{bold} correspond to the best fit value for each cluster.}

  \end{table}

\end{landscape}

\begin{landscape}

\begin{table}
  \caption[]{Globular clusters with 4 band photometry.}
\tiny
  \label{tableGC4}
  $$ 
  \begin{array}{ccccccccccccccccccccccccccc}
    \hline
    \hline
    \noalign{\smallskip}
 (1)  & (2)     & (3)     & (4)      & (5)       & (6)      & (7)       & (8)       & (9)      & (10)     & (11)      & (12)     & (13)      & (14)     & (15)     & (16)       & (17)       & (18)      & (19)      & (20)           & (21)           & (22)           & (23)           & (24)                & (25)                &(26)                 &        (27)         \\
 ID   & _{RA}   & _{DEC}  &  _{B}    & _{\sigma} &  _{V}    & _{\sigma} &  _{I}    & _{\sigma} &  _{U}    & _{\sigma} & _{FWHMB} & _{FWHMB}  & _{FWHMV}  & _{FWHMI}  & _{E(B-V)}  & _{E(B-V)}  & _{E(B-V)} & _{E(B-V)} & _{log(Age/yr)} & _{log(Age/yr)} & _{log(Age/yr)} & _{log(Age/yr)} & _{log(M/M_{\odot})} & _{log(M/M_{\odot})} & _{log(M/M_{\odot})} & _{log(M/M_{\odot})} \\
      & _{2000} & _{2000} & _{F435W} & _{F435W}  & _{F555W} &  _{F555W} & _{F814W} &  _{F814W} & _{F336W} &  _{F336W} & _{(pix)}  & _{(pix)} & _{(pix)} & _{(pix)} & _{Z=0.008} & _{Z=0.004} & _{Z=0.02} & _{Z=0.05} & _{Z=0.008}  & _{Z=0.004}  &  _{Z=0.02}  &   _{Z=0.05} &  _{Z=0.008}  &  _{Z=0.004}  &  _{Z=0.02}   &  _{Z=0.05}   \\
    \noalign{\smallskip}
    \hline
    \noalign{\smallskip}
29 & 0:14:12.89 &  -23:11:46.79 & 21.528 & 0.007 & 20.360 & 0.004 & 19.456 & 0.004 & 22.727 & 0.078 & 2.980  & 0.300 & 0.240 & 0.350 &  0.65 & 0.57 & \textbf{0.00} & 0.61 & 8.71 & 8.90 & \textbf{10.18} & 8.60 & 5.52 & 5.50 & \textbf{6.10} & 5.52 \\ 
30 & 0:14:14.82 &  -23:13:21.23 & 20.714 & 0.005 & 20.443 & 0.005 & 19.518 & 0.004 & 19.861 & 0.014 & 10.750 & 0.630 & 0.560 & 0.580 &  0.34 & 0.39 & 0.14 & \textbf{0.40} & 7.08 & 7.21 &  7.07 & \textbf{6.83} & 4.18 & 4.42 & 3.85 & \textbf{4.10} \\ 
31 & 0:14:04.22 &  -23:09:49.63 & 22.749 & 0.016 & 22.520 & 0.015 & 21.372 & 0.012 & 21.999 & 0.054 & 3.890  & 1.180 & 1.200 & 1.360 &  0.43 & 0.38 & 0.26 & \textbf{0.42} & 7.08 & 7.30 &  7.19 & \textbf{6.96} & 3.49 & 3.61 & 3.36 & \textbf{3.46} \\ 
32 & 0:14:01.28 &  -23:09:38.08 & 21.707 & 0.011 & 20.989 & 0.008 & 19.955 & 0.007 & 21.871 & 0.056 & 4.950  & 2.370 & 2.410 & 2.670 &  \textbf{0.00} & 0.00 & 0.02 & 0.00 & \textbf{9.20} & 9.35 &  9.05 & 8.96 & \textbf{4.97} & 5.05 & 4.95 & 4.93 \\ 
33 & 0:14:03.64 &  -23:10:44.86 & 21.981 & 0.011 & 21.358 & 0.009 & 20.355 & 0.009 & 22.250 & 0.059 & 3.270  & 0.790 & 0.790 & 0.890 &  \textbf{0.13} & 0.07 & 0.14 & 0.11 & \textbf{8.95} & 9.06 &  8.86 & 8.76 & \textbf{4.77} & 4.68 & 4.80 & 4.75 \\ 
34 & 0:14:03.91 &  -23:10:54.72 & 21.225 & 0.007 & 20.452 & 0.005 & 19.367 & 0.005 & 21.585 & 0.035 & 3.270  & 1.070 & 1.060 & 1.040 &  0.24 & 0.15 & \textbf{0.25} & 0.23 & 8.95 & 9.10 &  \textbf{8.86} & 8.73 & 5.26 & 5.14 & \textbf{5.29} & 5.23 \\ 
35 & 0:14:03.53 &  -23:10:48.80 & 22.263 & 0.012 & 21.643 & 0.009 & 20.716 & 0.010 & 22.087 & 0.061 & 3.490  & 1.020 & 1.030 & 1.040 &  0.49 & \textbf{0.68} & 0.45 & 0.94 & 7.75 & \textbf{6.91} &  7.81 & 6.60 & 4.29 & \textbf{3.79} & 4.32 & 3.99 \\ 
36 & 0:14:01.52 &  -23:10:06.25 & 22.226 & 0.012 & 21.564 & 0.009 & 20.542 & 0.008 & 22.367 & 0.078 & 3.080  & 0.860 & 0.960 & 0.960 &  0.56 & \textbf{0.00} & 0.00 & 0.00 & 7.99 & \textbf{9.26} &  9.05 & 8.92 & 4.57 & \textbf{4.75} & 4.70 & 4.68 \\ 
\noalign{\smallskip}
\hline
  \end{array}
  $$ 
For column descriptions see Table \ref{tableYoung}
%

  \end{table}

\end{landscape}

\onecolumn

\begin{table}
  \caption[]{Globular clusters with 3 band photometry.}
\tiny
  \label{tableGC3}
  $$ 
  \begin{array}{ccccccccccccc}
    \hline
    \hline
    \noalign{\smallskip}
 (1)      &     (2) & (3) & (4) & (5) & (6) & (7) & (8) & (9) & (10) & (11) & (12)&(13)  \\
 ID  & RA       & DEC     &  B      & \sigma &  V     & \sigma &  I    & \sigma & FWHMB      & FWHMB   & FWHMV & FWHMI   \\
   & 2000     & 2000    & F435W   & F435W  & F555W  &  F555W & F814W &  F814W & SExtractor~(pix) & ISHAPE~(pix) & ISHAPE~(pix) & ISHAPE~(pix) \\
    \noalign{\smallskip}
    \hline
    \noalign{\smallskip}

 37 &  0:14:08.84 &  -23:11:58.16 & 22.244 & 0.013 & 21.604 & 0.009 & 20.455 & 0.008 & 4.120 & 1.750 & 1.840 & 2.270 \\
 38 &  0:14:07.07 &  -23:10:33.77 & 22.456 & 0.017 & 22.199 & 0.015 & 21.554 & 0.016 & 5.730 & 3.150 & 2.690 & 2.140 \\
 39 &  0:14:05.77 &  -23:10:06.71 & 21.996 & 0.013 & 21.389 & 0.010 & 20.429 & 0.010 & 5.450 & 2.760 & 2.890 & 2.870 \\
 40 &  0:14:06.93 &  -23:10:47.15 & 20.602 & 0.006 & 19.978 & 0.005 & 19.035 & 0.005 & 5.120 & 1.530 & 1.580 & 1.600 \\
 41 &  0:14:04.97 &  -23:10:15.83 & 22.290 & 0.014 & 21.605 & 0.010 & 20.641 & 0.010 & 4.530 & 1.790 & 1.870 & 1.890 \\
 42 &  0:14:04.18 &  -23:10:34.09 & 21.671 & 0.008 & 20.997 & 0.007 & 19.985 & 0.006 & 3.640 & 1.040 & 1.120 & 1.280 \\
 43 &  0:14:04.94 &  -23:10:54.28 & 21.267 & 0.007 & 20.524 & 0.005 & 19.483 & 0.005 & 3.650 & 1.170 & 1.220 & 1.260 \\
 44 &  0:14:04.17 &  -23:10:53.98 & 21.295 & 0.008 & 20.560 & 0.006 & 19.383 & 0.005 & 3.970 & 1.610 & 1.550 & 1.540 \\
 45 &  0:14:04.67 &  -23:11:18.17 & 22.388 & 0.013 & 21.480 & 0.009 & 20.532 & 0.008 & 4.450 & 1.410 & 1.510 & 1.500 \\
 46 &  0:14:04.01 &  -23:11:06.07 & 22.566 & 0.012 & 21.793 & 0.010 & 20.784 & 0.010 & 3.200 & 0.750 & 0.940 & 0.930 \\
 47 &  0:14:04.40 &  -23:11:16.11 & 22.453 & 0.020 & 21.700 & 0.015 & 20.669 & 0.015 & 9.680 & 7.250 & 7.570 & 7.170 \\
 \noalign{\smallskip}
    \hline
  \end{array}
  $$ 
Column (1): Globular cluster ID. Column (2): RA. ~Column (3): DEC. ~Columns (4), (5), (6), (7), (8), (9): Photometry and the error 
in magnitude units.  ~Column (10): FWHM derived using SExtractor for the F435W band in pixels. ~Columns (11), (12), (13): FWHM derived using 
ISHAPE in pixels for each band.

  \end{table}

\end{document}